\documentstyle[twocolumn,prl,aps]{revtex}
\begin{document}   
\draft            
\title{ 
Zero-Temperature Phase Transitions of Antiferromagnetic Ising  
Model of General Spin on a Triangular Lattice} 
\author{Chen Zeng$^a$\cite{muetter} and Christopher L. Henley$^b$}  
\address{$^a$ Department of Physics, Syracuse University,
Syracuse, NY 13210, USA} 
\address{$^b$ Department of Physics, Cornell University,
Ithaca, NY 14853, USA} 
\date{\today}
\maketitle

\begin{abstract} 
We map the ground-state ensemble of antiferromagnetic Ising model
of spin-$S$ on a triangular lattice to an interface model whose  
entropic fluctuations are proposed to be described by an effective 
Gaussian free energy, which enables us to calculate the critical 
exponents of various operators in terms of the stiffness constant 
of the interface. Monte Carlo simulations for the ground-state
ensemble utilizing this interfacial representation are performed
to study both the dynamical and the static properties of the model. 
This method yields more accurate numerical results for the critical 
exponents. By varying the spin magnitude in the model, we find that 
the model exhibits three phases with a Kosterlitz-Thouless phase  
transition at ${3\over 2}<S_{KT}<2$ and a locking phase transition
at ${5\over 2} < S_{L} \leq 3$. The phase diagram at finite temperatures 
is also discussed.
\end{abstract}


\section{Introduction}
Over the years it has been found that there exist many two-dimensional 
classical spin models, discrete and continuous alike, whose ground-state 
manifolds are macroscopically degenerate and, more interestingly, also  
exhibit critical behaviours, i.e., spin-spin correlation functions 
within the ground-state ensembles decay with distance as power laws.  
The classification of universality class for these models has always 
been a challenging problem\cite{Liebmann}  

An earlier example of this kind is the antiferromagnetic Ising model 
on the triangular lattice. The exact solution for this model by 
Stephenson\cite{Stephenson} showed that although this model remains paramagnetic
at nonzero temperature, its ground state is critical. Later works 
by Bl\"ote {\sl et al} revealed yet another remarkable property of the  
ground-state ensemble of this model, namely, it permits a  
Solid-on-Solid (SOS) representation in which spin fluctuations 
are subsequently described by the fluctuating interface in the 
SOS model\cite{Blote}. Recent studies also demonstrated that this interfacial 
representation provides a valuable avenue for studying the  
ground-state ordering of quantum 
magnets\cite{quantum,Henley1}
and the ground-state roughness of oriented elastic manifolds in 
random media\cite{Zeng}.  Other recently studied models with  
critical ground states include three-state antiferromatic Potts  
model on the Kagom\'e lattice\cite{Huse,Chandra}, the $O(n)$ model 
on the honeycomb lattice\cite{Blote2,Kondev1}, the Four-Coloring model 
on the square lattice\cite{Kondev2,Kondev3}, and the square-lattice
non-crossing dimer model and dimer-loop model\cite{Raghavan}.  
On the other hand, some very similar models with degenerate 
ground states exhibit long-range order, such as the
constrained 4-state Potts antiferromagnet\cite{Burton}.

In this article we study the ground-state properties of antiferromagnetic 
Ising model of general spin on a triangular lattice which also belongs 
to the class of models mentioned above. Recent numerical studies of this  
model include Monte Carlo simulations\cite{Nagai,Honda} and transfer matrix 
calculations\cite{Lipowski}. Here we revisit this model by performing 
Monte Carlo simulations. The motivation of the present work is two-fold: 
(1)unlike previous simulations, we utilize the interfacial
representation directly in analyzing 
the simulation results, for example, we compute the stiffness constant 
of the fluctuating interface which, in turn, yields more accurate 
critical exponents of various operators; and (2) we also study the 
dynamical properties of this model for the first time making use of 
the interfacial representation.     

The body of the this paper is organized as follows. 
Section \ref{Model} describes the model Hamiltonian and maps it
onto a spin-1 problem whose interfacial representation is then 
described. In Section \ref{height-rep},
we propose an effective continuum theory for 
the long-wavelength fluctuations of the interface. Here we also  
show how to relate scaling dimensions of various operators 
to the stiffness constant of the interface, and derive some
other analytical results based on this ``height representation.''
This allows analytical understanding of the phase diagram
(Sec.~\ref{PhaseDiag}). 
Details of Monte Carlo simulations and numerical results 
on both dynamical and static 
properties are presented in Section \ref{MC-results}, 
including a comparison of the new and old approaches 
to determining the exponents.
As a conclusion, the paper is summarized and various possible extensions
are outlined, in Section \ref{Conc-Disc}.  

\section{The Model} 
\label{Model}

The antiferromagnetic Ising model of spin-$S$ on a triangular 
lattice can be described by the following Hamiltonian:
\begin{equation}
H = J \sum_ {{\bf r}} \sum_{{\bf e}} s({\bf r}) s({\bf r}+{\bf e}) 
\label{eq1}
\end{equation}
where the spin variable $s({\bf r})$ defined on lattice site 
${\bf r}$ of the triangular lattice can take any value from a 
discrete set $[-S, -S+1, \cdot\cdot\cdot, S-1, S]$, 
and the sum over ${\bf e}$ runs over three nearest-neighbor vectors 
${\bf e}_1$, ${\bf e}_2$ and ${\bf e}_3$ as shown in Fig.~\ref{fig1}.   
Here the coupling constant $J$ is positive
describing the antiferromagnetic exchange interaction between 
two nearest-neighbor spins: $s({\bf r})$ and $s({\bf r}+{\bf e})$. 

One important reason for interest in this model is that the 
$S\to \infty $ limit\cite{FNSinfty} 
is the same as the Ising limit of the
(classical or quantum) Heisenberg antiferromagnet on the triangular lattice
with Ising-like anisotropic exchange. 
That model was shown to exhibit a continuous classical ground state
degeneracy and unusual features of the selection by fluctuations
of ground states\cite{Heisenberg}.

The ground-state configurations of the above model given 
by Eq. (\ref{eq1}) consist of entirely of triangles on which 
one spin is $+S$, another is $-S$, and the third can be anything in $[-S,+S]$.
Thus, if spin $s(\bf r)$ takes an intermediate value $-S<s({\bf r})<S$, this
forces the six surrounding spins to alternate $+S$ and $-S$;
exactly which intermediate value $s(\bf r)$ takes does not matter in
determining whether a configuration is allowed. 

\subsection{Spin-1 mapping}

Therefore, this allows 
us to reduce each state $\{s({\bf r})\}$ to a state $\{\sigma({\bf r})\}$ 
of a {\sl spin-1} model, by mapping $s({\bf r})=+S$ to $\sigma({\bf r})=+1$, 
$s({\bf r})=-S$ to $\sigma({\bf r})=-1$, and intermediate values  
$-S<s({\bf r})<+S$ to $\sigma({\bf r})=0$. 
In this {\sl spin-1} representation of the model, 
the rules for allowed configurations are exactly the same as for
the $S=1$ model; however instead of being equal, the statistical weights 
have a factor $2S-1$ for each spin with $\sigma({\bf r})=0$. 
It should be noted that in the $S=1/2$ case, $s({\bf r})=\pm 1/2$
simply maps to $\sigma({\bf r})=\pm 1$. 

It can also be shown that the expectation of any polynomial
in $\{ s({\bf r}) \}$, 
in the ground-state ensemble of the spin-$S$ model, can be
written in terms of a similar expectation in the spin-1 model.
Specifically, one must simply replace
    \begin{equation}
      s({\bf r})^m \to \cases {S^m \sigma({\bf r}), &  m odd \cr
                              S^m [ (1-C_m(S)) \sigma({\bf r})^2 + C_m(S)], &  
                              m even \cr}
    \end {equation}
where (e.g.) $C_2(S) = {1\over 3} (1-S^{-1})$.
Thus there is no loss of information in this mapping. 
Indeed, in some sense, the extra freedom to have
$s({\bf r})$ vary from $-(S-1)$ to  $S-1$ is trivial:
once given that $s({\bf r})$ and $s({\bf r}')$ are 
intermediate spin values, there is no correlation between these values.

So we henceforth restrict ourselves 
to the spin-1 mapped model whose partition function for its ground-state 
ensemble can be written as: 
\begin{equation} 
Z=\sum_{\{\sigma(\bf r)\}} (2S-1)^{n_s} 
\;\; , 
\label{eq2}
\end{equation} 
where $n_s$ denotes the number of zero spins in a ground-state 
configuration ${\{\sigma(\bf r)\}}$. 
By varying the weight factor continuously in the spin-1 model,  
it would possible to give a precise meaning to {\it any} real value of $S$,
and to simulate such an ensemble. However, in this article we perform 
Monte Carlo simulations for an ensemble in which $2S$ takes only integer 
values. 

The spin-1 representation could be further reduced to a spin-1/2 
representation $\tilde \sigma({\bf r})$ as described
in Refs.~\onlinecite{Lipowski2,Lipowski,Honda}. 
They let 
   \begin{equation}
    \tilde \sigma ({\bf r}) \equiv \sigma({\bf r})+ k({\bf r})
    \label{spinhalf}
   \end{equation}
Here $k({\bf r})=0$ if $\sigma({\bf r})=\pm 1$ and 
if $\sigma({\bf r})=0$,  
$k({\bf r})=+1$ or $-1$ according to whether 
the surrounding spins are $(+1,-1,+1,-1,+1,-1)$ or the reverse.
Note this mapping is not invertible. 
The spin-$1/2$ representation is less satisfactory in that is 
arbitrarily breaks the up-down symmetry of correlation functions, 
but it was desirable for the transfer-matrix calculations
of Lipowski {\it et al}\cite{Lipowski}
since it reduced the number of degrees of freedom.

\subsection{Height mapping} 

We define a {\sl microscopic}, discrete-valued height function 
$z({\bf r})$ living on the vertex of the triangular lattice 
such that the step in $z({\bf r})$ between adjacent vertices 
is a function of the adjacent spins: 
\begin{equation} 
z({\bf r+e}) - z({\bf r}) = 
{1\over 2}+{3\over 2} \sigma({\bf r+e})\sigma({\bf r}) 
\;\; , 
\label{eq3} 
\end{equation} 
where $\sigma({\bf r})$ is the spin-1 operator and 
$\bf e$ can be any of the three nearest-neighbor vectors 
${\bf e}_{1,2,3}$. It is easy to show that the total 
change in height function, when traversed along any 
smallest loop, i.e, an elementary triangle, is zero. 
Therefore, $z({\bf r})$ is well-defined everywhere 
for the ground-state configurations, but it is not well-defined
in any excited state. 
This prescription generalizes
that originally introduced by
Bl\"ote et al for the case $S=1/2$\cite{Blote,Ogawa,FN1}
(the prescriptions agree in that case).

This type of height mapping differs from other sorts of mapping
(e.g. dualities) in a crucial way: since the spin microstates 
of the spin-1 model are mapped essentially one-to-one to
the height microstates, it is possible to perform Monte Carlo simulations 
and construct configurations $z({\bf r})$ after each time step.
We have found that analysis of the $z({\bf r})$ correlations is
much more efficient for extracting critical exponents than 
analysis of the spin correlations directly as was done in 
previous Monte Carlo simulations\cite{Nagai}.

\section{Height Representation Theory} 
\label{height-rep}

In this section we propose an effective
continuum theory which describes the
long-wavelength fluctuations of the interface. We also demonstrate 
how the critical exponents of various operators are determined 
by the stiffness constant of the interface.

\subsection{Effective free energy} 

To describe the interface in the rough phase, we must define a smooth
height field $h({\bf x})$ by coarse-graining the discrete field $z({\bf r})$.
As a first stage,  on every triangular plaquette formed by sites 
${\bf r}_1, {\bf r}_2, {\bf r}_3$, define a new discrete height
\begin{equation} 
h({\bf R}) \equiv {1\over 3} (z({\bf r}_1)+z({\bf r}_2)+z({\bf r}_3))
\;\; , 
\label{eq4} 
\end{equation} 
where ${\bf R}$ is the center of a triangle. The possible values of 
the $h({\bf R})$ are $\{ n/2 \}$, for any integer $n$.
(For the case $S=1/2$, the only possible values are integers.)
To each of these values corresponds a {\it unique}
ground-state spin configuration of the spin-1 model 
on that triangle, i.e., 
\begin{equation} 
s({\bf r})=\Phi_s( h({\bf r+u})-h_0({\bf r}) )
\;\; , 
\label{eq5} 
\end{equation} 
where ${\bf u}$ is any vector from a site to
the center of any adjoining triangle.
The mapping is many-to-one: the function $\Phi_s(h)$ has period 6.
Notice that the r.h.s. of Eq.(\ref{eq5}) turns out to be independent
of $\bf u$, but the periodic dependence on $h$ is phase-shifted by a
function $h_0({\bf r})$ which takes different 
values on each of the three $\sqrt3\times\sqrt3$ sublattices.
Essentially, we have mapped the $T=0$ ensemble of the
spin-1 problem into an equivalent interface problem.
Note that, given a configuration of $\{ h({\bf R}) \}$,
each $\sigma({\bf r})$ 
is specified (via Eq. (\ref{eq5})),
once for each adjoining triangle. 
The requirement that these six values of $\sigma({\bf r})$ 
coincide translates to a somewhat complicated set of contraints 
between pairs $h({\bf R})$ and $h({\bf R'})$ 
on adjoining triangles;
the difference $h({\bf R}) - h({\bf R'})$ 
may be 0, $\pm 1/2$, or $\pm 1$, 
but some of these are disallowed (depending on 
which $h()$ values are integer or half-odd-integer,
and on the orientation of ${\bf R}-{\bf R'}$). 
The weight of each configuration is given, as in (\ref{eq2}), by  
by $(2S-1)^{n_s}$. 

Fig.~\ref{fig1} shows the $h({\bf R})$ mapping explicitly where the 
spins $\sigma({\bf r})$ take values from $\{+1,0,-1\}$.  
The twelve states are arranged in a circle because the pattern 
repeats when $h\to h\pm6$. 

There are certain special ``flat states'' in which $h({\bf R})$ 
is uniform on all triangles. Each of these is periodic
with a $\sqrt3 \times \sqrt3$ unit cell -- in effect it
is a repeat of one of the triangles in Fig.~\ref{fig1}.
We shall name these states by writing the spins on the
three sublattices,``$(+,+, -)$'' and ``$(+,-,0)$'';
here ``$\pm$'' stands for $\sigma=\pm 1$. It should be 
noted that there are two non-equivalent species of flat state  
corresponding to integer, and half-integer valued $h({\bf R})$ 
respectively. They are non-equivalent in the sense that they are  
not related by {\sl lattice} symmetries. One of the species 
that is favored by the {\sl locking potential}
(see Eq. (\ref{eq-Vlock}) below)  
is what is previously called ``ideal'' 
states\cite{Kondev2,Kondev3,Raghavan,Burton}.  
   
Thus we can imagine that all states can be described
as domains of uniform $h({\bf R})$ separated by domain walls.
Finally, by coarse-graining $h({\bf R})$ over distances 
large compared to the lattice constant, one obtains $h({\bf x})$ 
which enters the conjectured continuum formula for 
the free energy, which is entropic in origin\cite{Blote},   
\begin{equation} 
F(\{ h({\bf x}) \} = \int d{\bf x}
\left\lbrack 
{K\over2} |\nabla h({\bf x})|^2 + V(h({\bf x}))
\right\rbrack  
\;\; ,  
\label{eq6} 
\end{equation} 
where $K$ is the stiffness constant of the fluctuating interface.

A lattice shift by one lattice constant
leaves the free energy invariant, but induces global shifts 
in height space $h({\bf x}) \to h({\bf x})\pm 1$; hence
the potential $V(\cdot)$ in (\ref{eq6}) must 
have period one.
It is typically approximated as 
  \begin{equation}
      V(h) \approx h_V \cos (2\pi h).
  \label{eq-Vlock}
  \end{equation}
Such a periodic potential, usually 
referred as the {\sl locking term}\cite{Jose}, favors the 
heights to take their discrete values one of the two types
of flat state, depending on the sign of $h_V$. 
For large $S$ we expect $h_V<0$, favoring the $(+,-,0)$ states,
in view of the large entropy of flippable spins; it is not so
sure which state is favored at smaller $S$, but this does not matter
for the critical exponents (see Sec.~\ref{Scaling} and~\ref{Operators}, below.

\subsection{Fluctuations and correlation functions}  
\label{Fluctuations}

In the {\sl rough phase}, by definition, 
the locking term is irrelevent, 
and so the long-wavelength fluctations 
of height variable $h({\bf x})$ 
are governed by the Gaussian  term of Eq. (\ref{eq6}):
\begin{equation} 
F(\{ h({\bf x}) \} = \int d{\bf x}
{K\over2} |\nabla h({\bf x})|^2  
=\sum_{\bf q} {K\over2} {\bf q}^2 |h({\bf q})|^2 
\;\; ,  
\label{eq7} 
\end{equation} 
where we have performed the Fourier transform. 
Hence by equipartition theorem, 
\begin{equation} 
S_h({\bf q}) \equiv \langle |h({\bf q})|^2 \rangle = {1\over {K {\bf q}^2}}  
\;\; . 
\label{eq8} 
\end{equation} 
Similarly, we can also measure 
the {\sl height-height difference function} in the real space as: 
\begin{eqnarray} 
C_h({\bf R}) 
& \equiv &  \frac{1}{2} \langle [h({\bf R})-h({\bf 0})]^2\rangle 
\nonumber \\ 
& = & \frac{1}{2\pi K} \ln(\pi R/a) + ... \;\; (R \gg 1) 
\;\; ,   
\label{eq9} 
\end{eqnarray}   
where $a$ is the lattice spacing cutoff.   

\subsection{Scaling dimensions}  
\label{Scaling}

Using Eq. (\ref{eq9}), we can compute the scaling dimension 
$x_O$ of any {\sl local} operator $O({\bf r})$, 
which is defined as in the correlation function, 
\begin{equation}
\langle O^*({\bf r)} O({\bf 0}) \rangle
\sim r^{-2x_O}
\;\; .
\label{eq10}
\end{equation}
By local operator, we mean that $O({\bf r})$ is a local  
function of spin operators in the vicinity of $\bf r$. 
Now, the same spin configuration is recovered when the height variable
$h({\bf R})$ is increased by 6.\cite{ferroJ2}
Thus any local operator $O({\bf r})$ 
is also a periodic function in the height space, and can consequently 
be expanded as a Fourier series: 
\begin{equation} 
O({\bf r}) = \sum_{G}  O_{G} e^{i G h({\bf r})}
\sim e^{i G_O h({\bf r})}
\;\; , 
\label{eq11} 
\end{equation} 
where $G$ runs over height-space reciprocal-lattice vectors
(i.e. multiples of $2\pi/6$). 
The last step of simplification in (\ref{eq11}) follows because
the scaling dimension $x_O$ of the operator $O({\bf r})$ 
is determined by the leading relevant operator in the above 
expansion, i.e., $G_O$ is the smallest $G$ with nonzero coefficient
in the sum.
Inserting Eq. (\ref{eq11}) into Eq. (\ref{eq10}) and making use 
of Eq.~({\ref{eq9}), we obtain the following: 
\begin{eqnarray}  
\langle O^*({\bf r)} O({\bf 0}) \rangle 
& = & \langle e^{-i G_O h({\bf r})} e^{i G_O h({\bf 0})} \rangle \nonumber \\ 
& = & e^{-G_O^2 C_h({\bf r})} \sim r^{-\eta_O}   
\;\; . 
\label{eq12}
\end{eqnarray} 
Therefore, the critical exponent $\eta_O$ associated with the 
operator $O({\bf r})$ is given by: 
\begin{equation} 
\eta_O \equiv 2 x_O = { 1 \over {2\pi K}} |G_O|^2
\;\; . 
\label{eq13} 
\end{equation}   

\subsection{Definition of operators} 
\label{Operators}

In this paper, besides the usual spin operator $\sigma({\bf r})$,
we also study the bond-energy operator $E({\bf r}+{\bf e}/2)$ 
for the reason that will become clear in the next section: 
\begin{equation} 
E({\bf r}+{\bf e}/2) =  
{1\over 2}+{3\over 2} \sigma({\bf r+e})\sigma({\bf r}) 
\;\; ,  
\label{eq14}
\end{equation} 
where ${\bf e}$ denotes one of the three nearest-neighbor vectors
as before. 

As discussed already, the spin operator on a given site 
has a periodicity of $6$
in the height space, from which a simple inspection shows 
that the bond-energy operator is also periodic in 
the height space with a periodicity of $3$. Therefore, 
the reciprocal lattice vectors of the most relevant operator 
in the Fourier expansion in Eq. (\ref{eq11}) are 
\begin{equation} 
G_{\sigma} = {2\pi\over6} , \;\;\;\;  G_{E} =  {2\pi\over3} 
\;\; , 
\label{eq15} 
\end{equation}  
for spin and bond-energy operators respectively. 

If a magnetic field is implemented by adding a term
$-H \sum _{\bf r} \sigma({\bf r}) $ to the Hamiltonian, 
then our dimensionless uniform ``magnetic field'' is defined 
by $H' \equiv H/T$. The exponents associated with 
$H'$ (and with the uniform magnetic susceptibility), 
are easily related to the correlation 
exponents of the uniform magnetization operator, 
\begin{equation}  
M({\bf R}) =  {1\over 3} 
( \sigma({\bf r}_1) + \sigma({\bf r}_2) +\sigma({\bf r}_3))    
\;\; , 
\label{eq-M} 
\end{equation} 
where ${\bf R}$ is the center of a triangle formed by sites 
${\bf r}_1, {\bf r}_2, {\bf r}_3$. A simple inspection of 
Fig.~\ref{fig1} shows that such an operator has a periodicity 
of $2$ in the height space, thus yielding: 
\begin{equation}  
G_{M} = {2\pi\over2} 
\;\; . 
\label{eq-GM} 
\end{equation}  

\subsection {Zone-corner singularities}
\label{Zone-corner}

Observe that the microscopic height variable
$z({\bf r})$ in any flat state is not uniform but is rapidly modulated  
with the wave vector ${\bf Q}={4\pi\over3}(1,0)$. The amplitude of modulation 
itself is a periodic function of the {\sl coarse-grained} height 
field $h({\bf x})$ which in turn implies 
that the correlation function decays with distance as a power-law, 
and consequently that its structure factor 
has a power-law singularity at ${\bf Q}$.

Such a zone-corner singularity is also directly connected to the 
singularity in the structure factor of the bond-energy operator. 
To see this, recall that there is a linear relation between the 
microscopic height variables and the bond-energy operator given 
by Eqs. (\ref{eq3}) and (\ref{eq14}), i.e., 
\begin{equation} 
E({\bf r}+{{\bf e}\over 2}) = z({\bf r}+{\bf e}) - z({\bf r})
\;\; . 
\label{eq-bond} 
\end{equation} 
Then it is interesting to note that the Fourier transform
$E_{\bf e} ({\bf q})$ of bond-energy operator given above 
turns out to be  
\begin{eqnarray} 
E_{\bf e} ({\bf q})
& \equiv & 
N^{-1/2} \sum_{\bf r} 
e^{ i{\bf q}\cdot({\bf r}+ {{\bf e}\over2})} E({\bf r}+{{\bf e}\over2}) 
\nonumber \\ 
&=&
-2i\sin ({1\over 2} {\bf q} \cdot {\bf e}) z({\bf q}) 
\;\; .  
\label{eq-Eq} 
\end{eqnarray} 
In other words, as a byproduct of measuring
$\langle |z({\bf q})|^2\rangle$, we have at the same time measured 
the structure factor of, say, the bond-energy operator of the same 
orientation specified by the nearest-neighbor vector ${\bf e}$:  
\begin{equation} 
S_E({\bf q}) \sim \langle |E_{\bf e}({\bf q})|^2\rangle 
= 4 \sin^2 ({1\over 2} {\bf q} \cdot {\bf e})  
\langle |z({\bf q})|^2\rangle
\;\; . 
\label{eq-SE} 
\end{equation} 
We will utilize this relation in Sec.~\ref{Structure} to extract 
the exponent of bond-energy operator from the Monte Carlo 
simulations.  

\subsection {Exact solution for $S=1/2$}

The $S=1/2$ triangular Ising antiferromagnet 
is exactly soluble, by the same techniques
which solve the ferromagnetic two-dimensional Ising model, 
and was immediately recognized to have critical behavior as $T\to 0$.
The spin and energy correlation functions were computed exactly by 
Stephenson; it transpires that $\eta_\sigma=1/2$ and $\eta_E=2$
exactly, implying through the arguments of Bl\"ote et al
(see Sec.~\ref{Scaling} and~\ref{Operators})
that the effective stiffness in Eq.~(\ref{eq7}) is $K= \pi/9$ exactly. 
The exponents implied by the interface scenario\cite{Blote}
-- in particular, the magnetic field exponent $\eta_M$ --
are fully confirmed by  numerical transfer-matrix 
computations.\cite{Blote3}

The Coulomb gas picture of Kondev {\em et al}\cite{Kondev4}, 
wherein the $S=1/2$ triangular Ising antiferromagnet 
is viewed as a fully-packed loop model\cite{Blote2} with fugacity 1, 
also predicts the exact exponents. 

\section {Phase Diagram}
\label {PhaseDiag}

In this section, we collect some consequences of 
the height representation for the phase diagram and the nature of 
the various phases within it.\cite{Chandra2}}

\subsection{Kosterlitz-Thouless and locking transitions}
\label{Locking}

The locking potential $V(\cdot)$ in (\ref{eq6}) favors the flat states. 
In view of (\ref{eq-Vlock}), its leading reciprocal-lattice vector 
is $G_V=2\pi$,
corresponding to a scaling index 
$ x_V = |G_V|^2 /{\pi K} = \pi/K $
for the corresponding conjugate field $h_V$.
It is well known that if $ 2 - x_V >0$, then $h_V$ becomes relevant
(under renormalization and the interface locks into one of 
the flat states.\cite{Jose} 
Since $K$ grows monotonically with $S$, 
such a locking transition occurs at a critical $S_L$ where
$K_L =\pi/2=1.57079...$\cite{Blote,Lipowski}.
In this ``smooth'' phase, any spin operator $O({\bf r})$
has long-range order, by arguments as in Sec.~\ref{Scaling}.

\subsection{Fluctuations in smooth phase}
\label{Smooth}

One of our aims in this paper was to pinpoint the
locking transition $S_L$, 
which demands that we have a criterion to distinguish these phases.
We must supplement Eq.~(\ref{eq8}), which shows the expected 
qualitative behavior of height fluctuations $\langle |h({\bf q})|^2\rangle$
in the rough phase, 
with a parallel understanding of the smooth phase.

In the smooth state, the symmetry (of height shifts) is broken
and a fully equilibrated system has long-range order,  such that
$\langle h({\bf x}) \rangle$ 
is well defined and uniform throughout the system. 
Fluctuations around this height, then, have at most short-range
(exponentially decaying) correlations.
Thus we expect them to have
a spatial ``white noise'' spectrum:
   \begin{equation}
      \langle |h({\bf q})|^2 \rangle \sim {\rm const}
   \label{eq-smooth}
   \end{equation}
for small $\bf q$. 

A phase with ``hidden'' order was suggested
by Lipowski and Horiguchi\cite{Lipowski,Lipowski2}. 
Numerical transfer-matrix calculations\cite{Lipowski} 
using the spin-1/2 representation indicated
$0 < \eta_{\sigma} <1/9$ for $2S>6$, which is impossible if the
spin correlations are derived from height fluctuations,\cite{Blote}
as we reviewed in Sec.~\ref{height-rep}.
An exotic phase to reconcile these facts
was to postulate a phase in which the interface was smooth and 
$\langle \tilde\sigma({\bf r}) \rangle\neq 0$,  
yet for the real spins $\langle \sigma({\bf r}) \rangle = 0$
as suggested by spin correlation measurements.

What does this imply for our height variable $h({\bf R})$, which
has a one-to-one correspondence with the real spin configuration
$\{ \sigma({\bf r}) \}$?
If the interface is smooth, then the probability distribution
of height values on a given plaquette, $P(h({\bf R}))$,
is well defined. 
In order to ``hide'' the order, it is necessary that $P(h)$
correspond to zero expectations of the spins.
Now, reversing $s({\bf r})$ on all three sites in the plaquette requires
$h \to h\pm 3$, as seen from Fig.~\ref{fig1}. 
One can convince oneself that,
to have ensemble average $\langle \sigma({\bf r})\rangle =0$, 
the distribution $P(h)$ must be at least as broad
as ${1\over 2} \delta (h-h_1) + {1\over 2} \delta (h-h_2)$, with
$h_1-h_2=\langle h \rangle \pm 3$, implying
the bound
\begin{equation}
{Var}[h({\bf R})] \equiv \langle h({\bf R})^2 \rangle
- \langle h({\bf R}) \rangle ^2 \ge (3/2)^2.
\label{eq-dhbound}
\end{equation}

\subsection{Finite temperature behavior}
\label{FiniteT}

At $T>0$, 
plaquettes with non-minimal energy are present and they
correspond to vortices in the function $h({\bf x})$.
Thus, unfortunately,  the height approach of analyzing 
simulations more or less breaks down. 
Nevertheless, one can still predict the $T>0$ phase diagram 
from knowledge of the $T=0$ stiffness constant derived from
our simulations. The shape of this phase diagram has already
been explained in Ref.~\cite{Lipowski}; here we note some
additional interesting behaviors which can be predicted
(following Ref.~\onlinecite{Blote}(b)) using the exponents associated with
vortices.

The other exponents in Kosterlitz-Thouless (KT)
theory are associated with elementary defects (often called vortices).
Indeed, it is easy to check (in this system) that 
the excess energy of a non-ground-state plaquette
is directly proportional to its vortex charge 
(a Burgers vector in height-space), so the effect of nonzero
temperature is simply to make the vortex fugacity nonzero.
The vortex exponent is $\eta_v= 1/\eta_\sigma$, so as usual
the vortex fugacity becomes relevant and defects unbind,
destroying the critical state, at the KT transition
defined by a spin exponent  taking the critical value
$\eta_{\sigma}=1/4$. 
If $\eta_{\sigma}>1/4$ at zero temperature, 
i.e. $K < K_{KT}\equiv 2\pi/9=0.69813...$,
then defects unbind as soon as $T>0$. 
Thus a zero-temperature KT transition occurs at
$S_{KT}$ defined by $K=K_{KT}$.\cite{Lipowski} 

Ref.~\onlinecite{Lipowski} did not, however, address the 
critical exponents of the correlation length $\xi(T)$
and the specific heat $C(T)$ as a function of temperature, which 
are also controlled by vortex exponents. 
Naively, if the energy cost creating one vortex is $E_c$, 
and if the minimum excitation is a vortex pair, 
then one would expect the low-temperature specific
heat to behave as 
       $C(T) \sim \exp (-2 E_c/T)$
and at $S=1/2$ this is indeed true\cite{Wannier}. 
However, the renormalization group\cite{Blote} 
shows the singular specific heat behaves as
   \begin{equation}
       f(T) \sim y(T)^{4/(4-\eta_v)}
   \end{equation}
where $y(T) = \exp (-E_c/T)$ is the vortex fugacity;
consequently when $\eta_v < 2$, the true behavior is
   \begin{equation}
       C(T) \sim \exp (-2 E_1/T)
   \end{equation}
with $E_1 = 2 E_c /(4-\eta_v) < E_c$. 
(Physically, part of the excitation energy is cancelled by
the large entropy due to the many places where the vortex pair
could be placed.)
This behavior has been observed in the 3-state Potts antiferromagnet
on the Kagom\'e lattice\cite{Huse}, and should occur
in the present system for all $S>1/2$. 

\subsection {Finite magnetic field}

It is interesting to consider the effect of a nonzero magnetic field $H'$.
It is known already that at $S=1/2$,\cite{Blote} 
such a field is an irrelevant perturbation, 
so that the system remains in a critical state, 
yet at sufficiently large $H$ it undergoes a
locking into a smooth phase,\cite{Blote3} 
approximated by any of the three symmetry-equivalent 
flat states of type ``$(+,+,-)$'' with magnetization $S/3$

As also already noted\cite{Lipowski},  
there is a critical value $S_{cH}$
defined by $\eta_\sigma(S_{cH})= 4/9$, 
beyond which $\eta_M = 9 \eta_\sigma < 4$
so that the system locks into long-range
order as soon as $H'$ is turned on. 
Within this regime, there are still two subregimes
with different behavior of $M(h)$ near $h=0$.
For $2 < \eta_M < 4$, the initial slope is zero, i.e., 
the susceptibility is not divergent; 
when $\eta_M < 2$, as occurs for $S \ge 2$, 
there is a divergent susceptibility and correspondingly
there should be a singularity at ${\bf q}=0$
in the spin structure factor 
$\langle |\sigma({\bf q})|^2 \rangle$. 

What do we expect in the locked phase at $S> S_{L}$?
Here the difference between the two kinds of flat states becomes crucial.
The $H'$ field favors the $(+,+,-)$ type of flat state, but entropy favors
the $(+,-,0)$ type of flat state. 
Thus we expect a transition to the $(+,+,-)$ state only at a nonzero 
critical field $H'_c$. 
On reducing $H'$ through $H'_c$, a twofold symmetry breaking occurs, in which 
one of the $+$ sublattices becomes the $0$ (disordered) sublattice; hence, 
this transition should be in the Ising universality class. 
Presumably the line $H'_c(S)$ meets the $H'=0$ axis at $S=S_{L}$.
There must also be line of locking transitions $S_{cH}(H')$, 
which terminates on the $H'=0$ axis at $S_{cH}$.

For $S=1/2$, the effect of the magnetic field was confirmed
numerically in Ref.~\onlinecite{Blote3}. 

\section{Monte Carlo Simulations and Results} 
\label{MC-results}

In this section we describe the implementation details 
of Monte Carlo simulations performed for spin-1 model 
in which $2S$ takes only integer values from $1$ to $8$.
We then present numerical results for the relaxation 
times of slow modes in the Monte Carlo dynamics. 
Two different methods of compute the critical exponents 
of the spin, bond-energy, and uniform-magnetization operators 
are described 
in different sub-sections: one in terms of the extrapolated 
stiffness constants of the interface and the other in 
terms of the singularities of the corrsponding structure 
factors. 

\subsection{Details of Monte Carlo Simulations}  

A spin is called {\sl flippable} if its six surrounding nearest-neighbor  
spins alternate between $+1$ and $-1$. Clearly, changing the 
value of this flippable spin results in another new spin configuration 
in the ground-state ensemble, provided that we start with a spin 
configuration in the ensemble. Moreover, such an update  
maintains the global tilt of the interface due to the 
{\sl local} nature of this update. This update will be used 
as our Monte Carlo update in this paper. Two slightly different  
cases arise for different values of $2S$: (1) for $2S=1$, 
the local update is precisely equivalent to a spin flip 
i.e., $\sigma({\bf r}) \rightarrow -\sigma({\bf r})$ due to 
the absence of zero spin; and (2) for all other values of $2S$, 
a random choice must be made in the local update: for 
example, $\sigma({\bf r})=0 \rightarrow \sigma({\bf r})=1$
or $-1$. (Recall $S$ denotes the spin magnitude of the 
original model.)

Let $n_s$ and $n_f$ denote the number of zero-spins and
flippable spins of configuration $\phi$. 
If an attempted single-spin update for $\phi$ results in
a new configuration $\phi^{\prime}$ with $n_s^{\prime}$ and $n_f^{\prime}$,
then the transition probability $W$ in accordance
with the detailed balance principle is:
\begin{equation}
W= W_0 \cdot min \{ 1, {n_f\over{n_f^{\prime}}} \} 
\cdot min\{1, (2S-1)^{n_s^{\prime}-{n_s}} \} 
\;\; ,
\label{eq16}
\end{equation}
where $W_0$ denotes the {\sl bare} transition probability:
$W_0={1\over n_f}$ for $2S=1$, and $W_0={1\over 2 n_f}$ for
$2S \ge 2$ which reflects the random choice to be made in 
the local update as discussed above. With the transition probability
given in Eq. (\ref{eq16}), it is straightforward to show that
the detailed balance principle is satisfied, i.e.,
$P(\phi) W(\phi \rightarrow \phi^{\prime})
=P(\phi^{\prime}) W(\phi^{\prime} \rightarrow \phi)$,
where $P(\phi)$ denotes the probability for
configuration $\phi$ to occur and $P(\phi) \sim (2S-1)^{n_s}$
since each spin configuration in the original spin-$S$ model 
has equal probability to occur.  
Note also that $n_f/n_f' = 1 + O(1/N)$ for large $N$, 
so  this rule is important only because of the finite system 
size. 

To implement in practice the transition probability given above,
we randomly select a site out of a list of the $n_f$ flippable sites, and
randomly update this spin to one of the two possible new spin values
if $2S\ge 2$ or simply flip this spin if $2S=1$. The total numbers
of zero spins $n_s^{\prime}$ and flippable spins $n_f^{\prime}$
in the resulting configuration are then computed. 
This update is subsequently accepted with a probability:
$min \{ 1, {n_f/{n_f^{\prime}}} \}\cdot
 min \{ 1, (2S-1)^{n_s^{\prime}-{n_s}} \}$.
A practical implementation of the transition probability given in
Eq. (\ref{eq16}) is thus achieved. 

Throughout this paper, a unit time or one Monte Carlo Sweep (MSC)
is defined such that there are $N_s$ attempts of updating within 
this unit of time (or one attempt per spin on average). Here $N_s$ 
denotes the total number of spins in the simulation cell. The 
simulation cell always contains $N_s=72 \times 72$ spins in this 
paper unless explicitly mentioned otherwise. Periodic boundary 
conditions are adopted. Since we always start with a flat state, 
the simulations are thus performed in the sector with a zero 
global tilt of the interface.

\subsection{Dynamical scaling: the relaxation time $\tau_{\bf q}$ } 
\label{Dynamical}

We now discuss the correlations between the configurations generated 
sequentially in the Monte Carlo simulations by studying the relaxation 
time of the slow modes in the model, namely, the Fourier modes 
$h_{\bf q}$ which play the role of an order parameter\cite{Henley1}.  
The linear-response dynamics of such a mode is usually formulated
as a Langevin equation, 
\begin{equation} 
{ {dh({\bf x},t)}\over{dt} } = 
-\Gamma { {\delta F(\{ h({\bf x}) \}) }\over{\delta h({\bf x})}}   
+\xi({\bf x}, t) 
\;\; , 
\label{eq17} 
\end{equation} 
where $\Gamma$ is the dissipation constant, and the static 
free energy functional $F(\{ h({\bf x}) \})$ is given by 
Eq. (\ref{eq6}). Here $\xi({\bf x}, t)$ is a stochastic 
noise generated in the Markov chain of Monte Carlo 
simulations. As it is expected that the correlation time 
of the slow mode under consideration is much longer than 
that of the noise, and since the update steps are local and independent,
it is proper to model $\xi({\bf x}, t)$ as 
Gaussian noise, uncorrelated in space or time: 
\begin{equation} 
\langle \xi({\bf x}, t) \xi({\bf x}^{\prime}, t^{\prime}) \rangle 
=2 \Gamma \delta ({\bf x} - {\bf x}^{\prime}) \delta (t- t^{\prime}) 
\;\; , 
\label{eq18} 
\end{equation} 
in which the choice of $2\Gamma$ ensures that the steady-state of the 
interface under the Langevin equation (\ref{eq17}) 
agrees with its equilibrium state 
under the free energy (\ref{eq6}).  

This linear stochastic differential equation can be solved easily by 
performing Fourier transform. Eq. (\ref{eq17}) thus reduces to 
\begin{equation} 
{ {dh({\bf q},t)}\over{dt} } = 
-\Gamma K |{\bf q}|^2 h({\bf q},t) +\xi({\bf q},t)
\;\; , 
\label{eq19} 
\end{equation} 
which implies an exponentially decaying correlation function of 
$\langle  h^{*}({\bf q},t)  h({\bf q},0) \rangle
\sim e^{-t/{\tau_{\bf q}}} $ with the relaxation time $\tau_{\bf q}$ 
given by 
\begin{equation}
\tau_{\bf q} = {1\over{\Gamma K}} |{\bf q}|^{-2} 
\;\; . 
\label{eq20} 
\end{equation} 
Therefore, the dynamical scaling exponent for the Monte Carlo dynamics, 
defined by $\tau_{\bf q} \sim  |{\bf q}|^{-z}$,
is always $z=2$ in the rough phase. 

To check this prediction on the dynamical scaling exponent 
in practice where the above continuum theory is regularized 
on a lattice, we compute the following auto-correlation function 
$C({\bf q},t)$ of the {\sl microscopic} height variable $z({\bf q})$:
\begin{equation}
C({\bf q},t) = 
{ 
{\langle z^*({\bf q},0) z({\bf q},t) \rangle 
 -|\langle z({\bf q},0)\rangle|^2} 
\over 
{\langle z^*({\bf q},0) z({\bf q},0) \rangle 
 -|\langle z({\bf q},0)\rangle|^2}
} 
\;\; ,  
\label{eq21}
\end{equation}
Here $\langle \rangle$ stands for the dynamical average, and the 
time $t$ is measured in unit of MCS. For each interger-valued 
$2S=1,2,...,8$, we perform $10^5$ MCS's with a flat initial configuration
and compute the auto-correlation functions upto $t \le 50$ for modes 
that correspond to the five smallest $|{\bf q}|^2$ values.  
In Fig.~\ref{fig2}, we display the results so obtained for $2S=1$.
Other cases of $2S$ are found to have very similar features.
It is clear from Fig.~\ref{fig2} that $\log_{10} C({\bf q},t)$
can be fitted very well by $a - t/\tau_{\bf q}$ where $a$ and the
relaxation time $\tau_{\bf q}$ are the fitting parameters. 
In other words, the relaxation is strictly exponential in all cases. 
Note that we used a cutoff $t=10$ in our fitting. The same fitting
procedure is carried out for other cases of $2S$.

The final results of the relaxation time $\tau_{\bf q}$ as a function 
of  $|{\bf q}|^2$ for $2S=1, ..., 6$ are shown in Fig.~\ref{fig3}; and for 
$2S=6,7,8$ as an insert. The fact that $\tau_{\bf q}$ scales as 
$|{\bf q}|^2$ for  $2S=1, ..., 5$ as indicated by the fitting in 
Fig.~\ref{fig3} thus shows that the ground-state ensembles for $2S=1, ..., 5$ 
are in the rough phase. On the other hand, it is indeed clear from 
the insert that for $2S=7$ and $8$, $\tau_{\bf q}$ curves downward as 
$|{\bf q}|^2 \rightarrow 0$ which is in sharp constrast to 
those of $2S=1, ..., 5$. From this, we conclude that ground-state 
ensembles for $2S=7$ and $8$ are in the flat phase. As for $2S=6$,
it is not conclusive from the data available whether $\tau_{\bf q}$ 
scales as  $|{\bf q}|^2$ or curves downward as $|{\bf q}|^2\rightarrow 0$.
Nonetheless, the fact that the relaxation time of the slowest mode
for $2S=6$ is longer than for any smaller {\it or larger} value of $S$,
suggests that $2S=6$ is very close 
to the locking transition. Further support for this phase diagram  
is also obtained by explicit calculations of stiffness constants 
and critical exponents which is discussed in the next section.       

\subsection {Stiffness constants and critical exponents}  
\label{Stiffness} 

As implied by Sec.~\ref{Fluctuations} , the stiffness constant of 
the fluctuating interface can be directly measured by studying 
the long-wavelength fluctuations of the height variables, i.e., 
their structure factor as given by Eq. (\ref{eq8}). It should   
be noted that concerning the task of calculating the Fourier 
components $h({\bf q})$ in Eq. (\ref{eq8}), it can be replaced 
by the approximation in terms of the {\sl microscopic} height 
variables $z({\bf q})$ given by   
\begin{equation} 
h({\bf q}) \approx z({\bf q}) \equiv {w_0\over\sqrt{N_s}}
\sum_{\bf r} e^{-i{\bf q}{\bf r}} z({\bf r})
\;\; , 
\label{eq22} 
\end{equation} 
where $\bf r$ labels a lattice site of the finite triangular
lattice of total $N_s$ lattice sites used in the simulation. 
Here $w_0=\sqrt{3}/2$ is the {\sl weight} of a lattice site,  
i.e., the area of its Voronoi region, which is introduced 
so that the {\sl microscopic} height variable $z({\bf q})$  
coincides with the {\sl coarse-grained} height variable 
$h({\bf q})$ in the long-wavelength limit (${\bf q} \rightarrow 0$).  
But unlike $h({\bf q})$, $z({\bf q})$ still contains features 
such as zone-corner singularities discussed in Sec.~\ref{Zone-corner} 
that are only manifested in miscroscopic height variables. 

Starting with a flat state, we perform $2\times 10^3$ 
MCS's as the equilibrium time; subsequent measurements 
of physical quantities are carried out at intervals of 
$20$ MCS's. This separation is a compromise between
the correlation times of small $\bf q$ modes and of larger $\bf q$ modes,
which are respectively much longer and somewhat shorter than 20 MCS
-- see Fig.~\ref{fig2}. Each run consisted of $8 \times 10^5$
MCS, i.e. $4\times 10^4$ measurements were taken;
these were subdivided into $20$ independent 
groups for the purpose of estimating statistical errors. 
The same procedure is used for all $2S=1,2,...,8$ reported  
in this paper. 

In Fig.~\ref{fig4}, we plot $\langle |z({\bf q})|^2\rangle ^{-1}$ vs. 
${\bf q}^{2}$ for $2S=1$, including all ${\bf q}$ in the first 
Brillouin zone. From the plot, we observe that
$\langle |z({\bf q})|^2\rangle ^{-1}$ is remarkably isotropic up  
to about ${\bf q}^{2} \sim 1.5$. This comes about because of 
the 6-fold rotational symmetry of the triangular lattice 
which ensures that {\sl anisotropy} occurs only in $q^6$  
and higher order terms, assuming that the function is analytic.
This is in constrast to other models 
defined on the square lattice where anisotropy already sets 
in at the order of $q^4$\cite{Raghavan,Kondev1}. The lower 
envelope of the data points in Fig.~\ref{fig4} corresponds to the line of 
$q_y=0$ in the $q$-vector space. Other cases of $2S$ are found 
to have very similar features as illustated in the insert of 
Fig.~\ref{fig4} where we plot the lower envelope for all $2S=1,2,...,8$.   
The structure factor of the height variables appears to 
diverge in the long-wavelength limit $|{\bf q}|^2
\rightarrow 0$ for all $S$ values, even 
for the largest $S$ values. 
(In the latter case, however, we believe one would
see the plot asymptote to a constant value, in a 
sufficiently large system; see below.)

Two other interesting features of the structure factor are also
revealed in the insert in Fig.~\ref{fig4}: (1) for $2S\ge 2$, it appears 
to indicate yet another singularity at the zone corner
${\bf q} \rightarrow {\bf Q} \equiv {4\pi\over 3}(1,0)$ 
in the thermodynamic limit $N_s\rightarrow \infty$; and (2) 
for $2S=1$, it approaches a constant instead. As already 
discussed in Sec.~\ref{Zone-corner}, the appearance of 
zone-corner singularities is expected, the precise nature of 
such singularities, however, is discussed in the next section.  
In the remaining of this section, we analyze the zone-center 
singularity to check if height variables behave as required by
Eq. (\ref{eq8}) for the rough phase and consequently extract
the stiffness constants. 

To further study the nature of zone-center singularity in terms 
of how $\langle |z({\bf q})|^2\rangle$ scales as a function of 
${\bf q}^{2}$ in the long-wavelength limit, we show the log-log  
plot of  $\langle |z({\bf q})|^2\rangle^{-1}$  vs. ${\bf q}^{2}$
for $2S=1,...,8$ in Fig.~\ref{fig5}. Comparing the simulation 
results for different systems sizes of $L=36$, $48$ and and $72$, 
we notice that data are well converged down to accessible 
small ${\bf q}$ vectors -- except for the case of $2S=6$ and $7$, 
where the finite size effect is still discernible. This is, of course, 
consistent with the fact that $2S=6$ and $7$ are close to the 
locking transition where the correlation length diverges;
it is interesting, however, to notice that their finite-size trends  
are different. 
In the case $2S=6$, the data plot for $L=72$
curves upwards less than that for $L=48$, while
in the case $2S=7$, the $L=72$ data show {\em more}
upwards curvature than the $L=48$ data.

By fitting $\langle |z({\bf q})|^2\rangle^{-1}$ to a function 
$q^{2\alpha}$ with $\alpha$ being the fitting parameter, we obtain, 
using the data of system size $L=72$ and a cutoff ${\bf q}^2 \le 0.5$, 
the exponent $\alpha=0.990(1), 0.988(1), 0.986(2), 0.984(2), 0.974(2)$ 
and $0.935(1)$ respectively for $2S=1, 2, 3, 4, 5$, and $6$. 
Apart from the case of $2S=6$, these values agree with 
$\alpha=1$ as in the predicted ${\bf q}^{-2}$ power-law singularity 
of the structure factor in the rough phase, Eq. (\ref{eq8}). 
As for $2S=7$ 
and $8$, $\langle |z({\bf q})|^2\rangle^{-1}$ clearly deviates from 
a power-law scaling and instead curves upwards to level off, which      
indicates that models with $2S=7$ and $8$ are in the smooth phases 
where $\langle |z({\bf q})|^2\rangle$ remains finite 
as ${\bf q} \to 0$, as discussed in Sec.~\ref{Smooth}.  
This conclusion is in excellent 
agreement with that inferred from dynamical scaling analysis presented
in Sec.~\ref{Dynamical}.   

It should be noted that in Fig.~\ref{fig5}, as a general procedure adopted  
throughout this paper in extracting numerical values of some physical 
quantities, we have averaged the data corresponding to the same magnitude 
of $|{\bf q}|^2$ to further reduce the effect due to statistical errors. 
The relative statistical error on each individual data point  
$\langle |z({\bf q})|^2\rangle$ of small ${\bf q}$, 
which is measured directly from the variance  
among the 20 groups, is found to range from $1\%$ to $3\%$. 
This is indeed consistent with the estimates of such relative 
errors from the relaxation times of the slowest modes of models 
with different values of $2S$ already given in Sec.~\ref{Dynamical}.  
It is perhaps also worth noting that another good check on  
the statistical errors on each data point is to compare the values 
of $\langle |z({\bf q})|^2\rangle$ for three ${\bf q}$ vectors
which are related by $120^\circ$ rotations in reciprocal space,
which ought to be equal by symmetry. For example, in the case of 
$2S=1$, the values of $\langle |z({\bf q})|^2\rangle$ for the 
three ${\bf q}$ vectors of the same smallest magnitude 
${\bf q}^2=0.0101539$ of system size $L=72$ are, respectively,
$285.551$, $280.528$, and $280.566$, from which one thus also 
obtain the relative error of about $1\%$. This observation therefore 
motivates the averaging procedure used in this paper. 

The stiffness constants can be subsequently determined by fitting
${\bf q}^{-2} \langle |z({\bf q})|^2\rangle^{-1}$ to the 
function $K + C_1 {\bf q}^2$ for the isotropic part of  
the data in which the stiffness constant $K$ and $C_1$ are the 
fitting parameters. The final fitting on the averaged data is shown
in Fig.~\ref{fig6} where we used a cutoff ${\bf q}^2 \le 0.5$ 
in the fitting. We also tried other different cutoffs of ${\bf q}^2\le 0.1$ 
and ${\bf q}^2 \le 1.0$, and found as expected that the stiffness  
is not sensitive to the value of cutoff as long as it falls into the 
isotropic part of the data. For example, we obtain, in the case of
$2S=1$, $K=0.3488\pm0.0022, 0.3490\pm0.0008$, and $0.3488\pm0.0006$ 
for cutoff ${\bf q}^2 \le 0.1, 0.5$, and $1.0$ respectively. 
Therefore, taking into account of the uncertainty introduced 
due to the cutoff, our final estimate for the stiffness constant 
is then $K=0.349\pm0.001$ which is in excellent agreement 
with the exact value $K_{\mbox{exact}}=0.349065...$.  
Similar procedure is carried out for other cases of $2S$ and 
the results are tabulated in Table I. In the same table, we also 
give the value for the critical exponents of spin, bond-energy 
and uniform magnetization operators which are obtained   
straightforwardly according to Eqs. (\ref{eq13}), (\ref{eq15}) 
and  (\ref{eq-GM}). 
The agreement of our $\eta_\sigma^{(K)}$ values 
with the ``$\eta_\sigma$'' values from transfer-matrix eigenvalues 
(see Table~I of Ref.~\onlinecite{Lipowski}, 
is quite close and becomes better as $S$ grows (until $2S=6$.) 

As already discussed in Sec.~\ref{FiniteT}, 
a Kosterlitz-Thouless (KT) transition occurs at a critical value $S_{KT}$
where $\eta_{\sigma}=1/4$, such that for  $S>S_{KT}$ algebraic correlations
persist even at small finite temperatures.
It is clear from our data that $S_{KT}>3/2$.

As for $2S=6$, the value of ${\bf q}^{-2} \langle |z({\bf q})|^2\rangle^{-1}
=1.75\pm 0.06$ at the smallest nonzero ${\bf q}^2=0.010153$
is already larger than $K_L=\pi/2=1.57079$. That is, even if
the system may have a ``rough'' behavior at the length scales probed in 
the simulation, the stiffness constant
is such that the locking potential is relevant and must dominates 
at sufficiently large length scales, as discussed in Sec.~\ref{Locking}.
A similar observation has already been used to argue that the constrained
Potts antiferromagnet is in a smooth phase\cite{Burton}.
This fact together with the poor fitting using the formula suitable  
for the rough phase (see the top curve of Fig.~\ref{fig6})
leave us little choice but to conclude that the ground-state ensemble  
for $2S=6$ also falls into the smooth phase, or possibly is
exactly at the locking transition.  

Just as the finite-size effect for $2S=6$ was severe both for the
spin-spin correlations (measured via Monte Carlo\cite{Nagai,Honda})
and also in spin-operator eigenvalues 
(measured via tranfer-matrix,\cite{Lipowski})
we similarly find it is severe for height fluctuations.
However, in view of the exponential relationship between 
the exponents and the stiffness constant, the latter measurements
are much more decisive as to the true phase of the system.

To sum up, based on the analysis on the nature of the singularity 
in the height structure factor at the long-wavelength limit and 
the numerical results on the stiffness constants, we thus conclude 
that the model exhibits three phases with a KT phase  
transition at ${3\over 2}<S_{KT}<2$ and a locking phase transition
at ${5\over 2} < S_{L} \le 3$. 

\subsection{Structure factor and zone-corner singularity}  
\label{Structure}

Another more traditional approach\cite{Nagai} in calculating the critical
exponents of various operators is to compute the corresponding 
structure factors and analyze the power-law singularities at the 
appropriate ordering wave vectors. Namely, if the correlation function 
of an operator $O$ decays with distance as power-law (thus critical)
\begin{equation} 
\langle O({\bf r}) O({\bf 0}) \rangle \sim 
{  {e^{i{\bf Q}\cdot{\bf r}}}\over r^{\eta_O} } 
\;\; ,  
\label{eq23} 
\end{equation} 
then its structure factor near the ordering vector ${\bf Q}$ shows a 
power-law singularity   
\begin{equation} 
S_O({\bf q=Q+k}) \sim {\bf k}^{2(x_O-1)} 
\;\; ,  
\label{eq24} 
\end{equation} 
from which the critical exponent $\eta_O \equiv 2x_O$ can be 
numerically extracted. Here in this section, we adopt this 
approach to calculate the critical exponents of spin,
bond-energy, and uniform-magnetization
operators so as to compare with those obtained 
from the stiffness constant.   

As given by Eq. (\ref{eq-SE}),    
$S_E({\bf q=Q+k}) \sim \langle |z({\bf q=Q+k})|^2\rangle$.
Here ${\bf Q}={4\pi\over3}(1,0)$ is the ordering vector of 
the bond-energy operator. Therefore the interesting feature
of structure factor of height variables, namely, the appearance 
of zone-corner singularity as shown in Fig.~\ref{fig4}, 
is not only expected but also very useful in extracting the critical 
exponent $\eta_E$. 

Of course, such a zone-corner singularity 
can also be understood within the framework of interfacial 
representation, as in Sec.~\ref{height-rep}, particularly 
Subsec.~\ref{Zone-corner}.
(Similar zone-corner singularities have been studied in
Refs.~\onlinecite{Kondev2} and \onlinecite{Raghavan}.)
Finally, according to the exact result $\eta_E=2$ ($x_E=1$) in the case of 
$2S=1$, i.e., $S_E({\bf q=Q+k}) \sim {\bf k}^{2(x_E-1)} \rightarrow 
const.$, the puzzling absence of the zone-corner singularity
for $2S=1$ as shown in Fig.~\ref{fig4} is also resolved. 

In Fig.~\ref{fig7}, we plot $\log_{10}S_E({\bf q})$
vs. $\log_{10}|{\bf q-Q}|^2$ 
where we have averaged data points with the same magnitude of 
$|{\bf q-Q}|^2$. Fitting $S_E({\bf q})$ to the function 
$|{\bf q-Q}|^{2(x_E-1)}(C_1+C_2 |{\bf q-Q}|)$ where $x_E, C_1$ and $C_2$ 
are the fitting parameters, we obtain the critical exponents 
$\eta^{(S)}_E$ which are tabulated in Table I.
In practice, we used two different 
cutoffs in the fitting: $|{\bf q-Q}|^2 \le 0.1$ and $\le 0.5$. The 
fitting for the latter is shown in Fig.~\ref{fig7}, and the final quoted 
errors take into account the uncertainty due to the cutoffs. 

Similarly, we also computed the structure factor for the spin 
operator $S_{\sigma}(\bf q)$ using Fast Fourier transform while 
computing the height-height correlation function within the same  
Monte Carlo simulations.    
Results are shown in Fig.~\ref{fig8} and the extracted exponents are also 
tabulated in Table I. Fitting precedure used is exactly the same 
as that for bond-energy except that we fit $S_{\sigma}({\bf q})$ 
to the function $C_1 |{\bf q-Q}|^{2(x_{\sigma}-1)}$ with $C_1$ and 
$x_{\sigma}$ being the fitting parameters. From Table I,
we note that the critical exponents extracted in this way 
are in good agreement with those obtained from stiffness 
constant utilizing the interfacial representation, however, 
the latter yields much better statistical errors by an order of 
magnitude using the same Monte Carlo simulation data.  
This clearly demonstates the superiority of the interfacial 
representation in extracting critical exponents from numerical 
data. Similar points were made regarding other models, but based on  
much less extensive simulation data, in Refs.~\onlinecite{Kondev2}
and \onlinecite{Raghavan}. 

Similar fits were attempted for $2S=6$,
yielding $\eta^{(S)}_E (2S=6) = 0.53\pm0.41$ and 
$\eta^{(S)}_\sigma (2S=6)= 0.236\pm0.036$. 
While the statistical error on $\eta^{(S)}_E (2S=6)$ 
is too large to render the fitting meaningful, the increase in 
the value of $\eta^{(S)}_\sigma (2S=6)$ when compared with 
$\eta^{(S)}_\sigma (2S=5)$ is added evidence that $2S=6$ is {\sl not}
in the rough phase; if it were still rough at that
value of $S$, we would have expected a continuation of
the decreasing trend of $\eta^{(S)}_\sigma$ with $S$.

As for the cases of $2S=7$ and $8$, 
the structure factors of both the spin
and bond-energy operators
show {\it weaker} than power-law behavior as ${\bf q} \to {\bf Q}$, as in 
Figs.~\ref{fig7} and~\ref{fig8},
but they increase to a larger value (not seen in these logarithmic plots)
right {\it at} $\bf Q$.
This is indeed consistent with the 
$\delta$-function singularity. 
expected if these cases fall into the smooth phase 
with long-ranged order of the spin and bond-energy operators.

Finally, we consider the uniform magnetization correlation exponent $\eta_M$.
When $S>3/2$, it can be predicted (see $\eta^{(K)}_M$ in Table I) that 
$\eta_M< 2$, implying a divergent (ferromagnetic)
susceptibility and a divergent structure factor $S_M({\bf q})$ 
as ${\bf q}\to 0$
Now, due to the linear relation (\ref{eq-M})
between $\{ M({\bf R}) \}$ and $\{ \sigma({\bf r}) \}$, 
we immediately obtain 
$S_M({\bf q}) \sim S_\sigma({\bf q})$  near ${\bf q}=0$, 
just as $S_E({\bf q}) \sim \langle |z({\bf q})|^2\rangle$
near ${\bf q}={\bf Q}$
(see Sec.~\ref{Zone-corner} and Eq.~(\ref{eq-SE}))
Thus, a singularity at ${\bf q}=0$ 
is expected in the structure factor of spin operator which is plotted in 
Fig.~\ref{fig9}. From this figure, it appears 
that only for $2S=4$, $5$, and $6$ does $S_M({\bf q})$ show a
power-law singularity indicated by a straight line in this
log-log plot. This confirms the prediction based
on the stiffness constant; however, 
the numerical values of $\eta_M$
extracted this way (see Table I) differ considerably from those
calculated from the stiffness constant in the case of
$2S=5$ and $6$. 

It is also apparent from Table I that $\eta_\sigma^{(S)}$ is 
systematically overestimated as compared with the more accurate
value derived from height fluctuations.
We suspect that a similar overestimation affected the
values of $\eta_\sigma$  that
were deduced from the finite-size scaling of the 
susceptibility of the staggered magnetization\cite{Nagai,Honda}
(this obviously measures 
the same fluctuations seen in $S_\sigma({\bf q})$ near $\bf Q$.)
Those data (also quoted in Ref.~\onlinecite{Lipowski})
have quoted errors about four times as large as ours
for $\eta_\sigma^{(K)}$.
Their exponent values are all noticeably larger than the 
accurate value ($\eta_\sigma^{(K)}$ or $\eta_\infty$ from 
Ref.~\onlinecite{Lipowski}) -- becoming {\it worse} as $S$ grows
(for $2S=4,5$  the difference is twice their their quoted error.) 
Clearly the systematic contribution to their errors was underestimated.
The transfer-matrix method\cite{Lipowski}
ought to provide the 
effective exponent $\eta_\sigma$ for spin correlations
on length scales comparable to the strip width, and hence
is likewise expected to overestimate $\eta_\sigma$;
indeed, every $\eta_\sigma$ value found in Ref.~\onlinecite{Lipowski}
is slightly larger than our corresponding $\eta_\sigma^{(K)}$ value.

\subsection{Smooth Phase}   
\label{MC-Smooth} 

Which type of flat state is actually selected in the smooth phase?
Fig.~\ref{fig10} shows the measured expectation of
$n_s$, the number of zero spins in the spin-1 representation,
for $1 \leq 2 S \leq 8$. 
As $S$ grows, it is found that $\langle n_s\rangle$ approaches its 
maximum allowed value $N_s/3$ as in the $(+,-,0)$ state,  
rather than zero, as in the $(+,+,-)$ state.
Thus, the flat states with 
half-integer valued $h({\bf R})$ in Fig.~\ref{fig1} are being selected
in the smooth phase.
Translating back to the spin-$S$ model, this means
that spins on two sublattices of the triangular lattice 
take the extremal values,  $+S$ and $-S$ respectively,
while spins on the third sublattice remain disordered.

It is perhaps more illuminating to study the distribution of height 
variables to probe the height fluctuations in the smooth phase. 
To this end, we also show, in Fig.~\ref{fig10}, the histogram of  
height variable $h({\bf R})$ in the cases of $2S=2$ and $2S=8$, 
which is measured for a {\sl typical} configuration generated in 
the Monte Carlo simulations.\cite {FN-fig10}. 
The broad distribution 
observed in the case of $2S=2$ ($S<S_L$) 
evolves to a narrowly peaked distribution 
in the case of $2S=8$ ($S>S_L$). (It decays as
$\exp(-{\rm const}|h-\langle h \rangle|)$.) This supports the   
intuitive picture presented in Sec.~\ref{Smooth}. 
Furthermore, the center of this peaked distribution is half-integer 
valued. (Numerically, the mean is $\langle h\rangle =0.46$ for 
the distribution plotted in Fig.~\ref{fig10}.)
In other words, the locking potential $V(h)$ favors the $(+,0,-)$
type of flat state, in which one sublattice is flippable,
rather than the $(+,+,-)$ type of flat state. (See Fig.~\ref{fig1}).
This kind of flat state was also expected 
analytically in the limit of large $S$
\cite{Horiguchi2,Horiguchi}. 

We have also computed ${\rm Var} (h)$ for each value of $S$, in two ways.
First, ${\rm Var}(z)$ is just normalization factors times
$\sum _{\bf q \neq 0} \langle |z({\bf q})|^2 \rangle$, which we accumulated
throughout the Monte Carlo run, as described earlier in this 
section; then it can be shown that
$Var(h) = Var(z) -{1\over 3} + {1 \over 2} \langle n_s \rangle$ exactly.
For $N_s=72$ this gives ${\rm Var}(h) = 1.06$ and $0.20$ 
for $2S=2$ and $2S=8$, respectively, showing the contrast of 
the rough and smooth behavior.
Secondly, we can compute ${\rm Var}(h)$ directly from the histogram
(from one snapshot) seen in Fig.~\ref{fig10}; this gives respective
values $1.1$ and $0.15$, in satisfactory agreement with the first method.

The exotic ``hidden order'' phase\cite{Lipowski,Lipowski2}
(see Sec.~\ref{Smooth})
can be ruled out on the basis of these data:
according to Eq.~(\ref{eq-dhbound})
the variance of $h({\bf R})$ should be at least $(3/2)^2=2.25$
in the hidden-order phase,
while our measurements indicate it is at most only $0.20$.
Furthermore, for $2S=7$ and $8$, the structure factor $S_\sigma({\bf Q})$ 
at the zone-corner wave vector $\bf Q$ 
(not plotted) was much larger than at nearby $\bf q$;
that direct suggests  a $\delta$-function 
singularity in the thermodynamic limit, i.e., existence of
long-ranged spin order in which 
$\langle s({\bf r})\rangle \ne 0$ on at least two of the sublattices.

Additionally, the spin structure factor $S_\sigma({\bf q})$ 
near the zone-corner wave vector $\bf Q$ (Fig.~\ref{fig8})
showed a striking curvature in the ``smooth'' cases $2S=7$ and $8$,
quite different from the behavior at smaller $S$. This makes it plausible that 
$S_\sigma({\bf q}) \to {\rm constant}$, so that
spin fluctuations have short-range rather than power-law correlations
for $S>S_L$.  
(It was not emphasized in Ref.~\onlinecite{Lipowski},
but power-law correlations are implied if one takes seriously 
their measured values $0< \eta_\sigma < 1/9$ for $2S=7,8$.)

We propose, then, that actually $\eta_{\sigma} = \eta_{E} = \eta_{M} =0$
for $S> S_{c2}$, as in the simplest picture of the smooth phase,
and that the observed nonzero values are simply finite-size effects
due to the very slow crossover from rough to smooth behavior near
a roughening transition 
(see Sec.~\ref{Disc-crossover}, below,  for further discussion.)

\section{Conclusion and Discussion} 
\label{Conc-Disc}

To conclude, in this article, we have investigated the 
ground-state properties of the antiferromagnetic 
Ising model of general spin on the triangular lattice 
by performing Monte Carlo simulations.   
Utilizing the interfacial representation, we extrapolated 
the stiffness constants by studying the long-wavelength 
singularity in the height variables, which in turn lead 
to straightforward calculation of critical exponents 
of various operators within the framework of height 
representation. The results so obtained are further 
compared with those extracted from a more tranditional   
method, and demonstrate that the method in terms of 
height representation method is by far the preferable
one for extracting the critical exponents.  
We also analyzed both the dynamical and static properties 
of the model in order to map out the phase diagram which consists of
three phases with a Kosterlitz-Thouless phase
transition at ${3\over 2}<S_{KT}<2$ and a locking phase transition
at ${5\over 2} < S_{L} \le 3$.   
Even in the smooth state, 
analysis of the height fluctuations (as in ${\rm Var}(h)$
was helpful in resolving questions which are made difficult by
the strong finite-size effects near the locking transition.

\subsection{Rational exponents?}

One of our initial motivations for this study was the possibility 
of finding rational exponents even for $S>1/2$.
We believe the results in Table~I are the first 
which are accurate enough to rule out this possibility.
Indeed, $\eta_\sigma(2S=4)\approx 3/16$ and $\eta_\sigma(2S=5)\approx4/27$,
with differences similar to the error (0.001). 
But {\it any} random number differs from a rational number with 
denominator $<30$ by the same typical error.
The exception is that $\eta_\sigma^{(K)}(2S=6)$ is quite close to $1/9$,
but we have given other reasons to be suspicious of this value.

\subsection {What is $S_{L}$?}
\label{Disc-crossover}

Another intriguing question was whether
the critical values $2S_{KT}$ and $2 S_{L}$ are exactly integers.
Previous data\cite{Lipowski} suggested that
$S_L\equiv 3$ exactly, and had large enough errors
that $S_{KT}=3/2$ could not be excluded.
Since $\eta_\sigma(S_{KT})\equiv 1/4$ and $\eta_\sigma(S_L)\equiv 1/9$, 
this question was answered by the preceding subsection:
we find that definitely $S_{KT}<3/2$. Furthermore, we suspect
$S_{L} < 3$ as concluded in Sec.~\ref{Stiffness}
since the effective stiffness at the length scale we access
is more than enough to drive the system to the locked phase.

The question of the value of $S_L$ suggests paying closer
attention to the behavior of systems near the locking transition.
It has been noted previously how the locked phase tends to 
behave qualitatively like the rough phase in a finite-size system, since the
crossover is a very slow function of size.\cite{Blote3}
This is consistent with the apparent power-law behaviors observed
at $S>S_{L}$ in previous studies\cite{Nagai,Lipowski} and with 
the tendency of those studies to overestimate the exponents $\eta_{\sigma}$
and $\eta_{E}$ (as compared with our more accurate estimates.)
This would suggest that, if extensive finite-size
corrections were included in our analysis, they would 
reduce our estimate of $S_{L}$ a bit further, i.e. 
we would more definitely conclude that $2S=6$ is in the locked phase.

Our analysis near the locking transition at $S_{L}$ suffers from
our ignorance of the expected functional form of the critical behavior
as a function of $S-S_{L}$.  
A study of the roughening transition\cite{Evertz}
used the Kosterlitz-Thouless (KT) renormalization group to 
derive analytic approximations for the total height fluctuation
(closely analogous to  ${\rm Var}(h)$
in our problem), which made it possible to overcome very strong 
finite-size effects and fit the roughening temperature precisely.
Use of KT finite-size corrections was also essential in 
extracting meaningful numbers from transfer-matrix calculations
near the locking transition induced by a magnetic field in 
Ref.~\onlinecite{Blote3}.
Thus, a similar adaptation of the
KT renormalization group to give expressions
for the behavior of $\langle | z({\bf q} |^2 \rangle $, as a function
of (small) $|{\bf q}|$ and $S-S_{L}$,  or the functional form of $K(S)$ near
$S_{L}$, could make possible a 
more conclusive answer as to whether $S_{L}=3$ exactly. 

\subsection{Possible improved algorithms}

Since the long-wavelength behavior in this model (in its rough phase) is
purely Gaussian with $z=2$ (see Sec.~\ref{Dynamical}), the critical slowing
down is particularly transparent. 
It seems feasible to take advantage of the existence of a height
representation to develop an acceleration algorithm. 
For example, it might be possible to extend the cluster algorithms 
which are known for the $S=1/2$ 
triangular Ising antiferromagnet.\cite{cluster}
These are well-defined at $T>0$, but their effectiveness seems 
to depend in a hidden fashion on the existence of the height representation
when $T\to 0$. 

An intriguing alternative approach starts from the observation that
at long wavelengths the system obeys Langevin dynamics 
(see Sec.~\ref{Dynamical}
and Ref.~\onlinecite{Henley1}). 
Fourier  acceleration\cite{Batrouni}, a nonlocal algorithm
(efficient owing to use of the Fast Fourier Transform algorithm),
is known to be effective in such cases.
The key is to replace the uncorrelated noise function $\xi({\bf x},t)$
of Eq.~(\ref{eq18}) with a new correlated noise function having 
$\langle |\xi ({\bf q},t)|^2\rangle \sim 1/|{\bf q}|^2$. 
This might be implemented by first constructing a random function
with such correlations,  and then updating flippable spins
with probabilities determined by that function, in such a fashion as
to satisfy detailed balance.

Additionally, it may be possible to analyze transfer-matrices
using the height representation.
Quite possibly this would yield
an order-of-magnitude improvement in the accuracy 
of the numerical results, for the same size system,
similar to the improvement in analysis of Monte Carlo data.
The transfer matrix breaks up into sectors corresponding to the step 
made by $z({\bf r})$ upon following a loop transverse to the 
strip (across the periodic boundary conditions.
Then the stiffness can be extracted directly from
the ratio of the dominant eigenvalues of two such sectors;
such an analysis is already standard for
quasicrystal random tilings, for which the long-wavelength degree of
freedom is also an effective interface
\cite{quasicrystal}.

\acknowledgements
C.Z. gratefully acknowledges the support  
from NSF grant DMR-9419257 at Syracuse University.
C.L.H. was supported by NSF grant DMR-9214943.

\newpage 
\widetext 

\begin{table}
\caption{Stiffness constant and critical exponents. Here  
$\eta_{\sigma}^{(K)}$, $\eta_{E}^{(K)}$ and $\eta_{M}^{(K)}$
are the estimates for the critical exponents of spin and
bond-energy operators calculated from the stiffness 
constant $K$ as done in Sec. V(C), while $\eta_{\sigma}^{(S)}$,  
$\eta_{E}^{(S)}$, and $\eta_{M}^{(S)}$ stand for the same critical 
exponents, but extracted from the singularities of 
their respective structure factors in Sec. V(D). Estimated 
errors are also given in the parenthesis.}   
\begin{tabular}{|c|c|c|c|c|c|c|c|} \hline
$2S$  & $K$ & $\eta_{\sigma}^{(K)}$ & $\eta_{E}^{(K)}$ & $\eta_{M}^{(K)}$   
            & $\eta_{\sigma}^{(S)}$ & $\eta_{E}^{(S)}$ & $\eta_{M}^{(S)}$ 
\\ \hline\hline 
1 & 0.349(0.001) & 0.500(0.002) & 2.001(0.008) & 4.502(0.018) 
                 & 0.511(0.013) & 1.844(0.057) &             \\ \hline
2 & 0.554(0.003) & 0.315(0.001) & 1.260(0.006) & 2.836(0.013)
                 & 0.332(0.016) & 1.340(0.072) &             \\ \hline 
3 & 0.743(0.004) & 0.235(0.001) & 0.940(0.005) & 2.114(0.011) 
                 & 0.254(0.019) & 1.047(0.082) &             \\ \hline 
4 & 0.941(0.006) & 0.186(0.001) & 0.742(0.004) & 1.670(0.010)
                 & 0.203(0.022) & 0.791(0.092) & 1.634(0.014)\\ \hline
5 & 1.188(0.008) & 0.147(0.001) & 0.588(0.004) & 1.322(0.009)
                 & 0.180(0.026) & 0.504(0.115) & 1.560(0.015)\\ \hline
6 & 1.597(0.015) & 0.109(0.001) & 0.437(0.004) & 0.984(0.009)
                 & 0.236(0.036) & 0.530(0.410) & 1.527(0.016)\\  \hline\hline   
\end{tabular}
\end{table} 

\twocolumn 

\begin{figure}
\caption{Twelve flat states of the ground-state ensemble. 
Each flat state is simply specified by its spins on three sublattices
A, B, and C of the triangular lattice since all spins on same 
sublattice take the same value. The height variable 
$h({\bf R})$ defined at the center of an elementary triangle  
according to Eq. (4), which is uniform for each of these twelve
states, is also shown. Note that $h({\bf R}) \rightarrow 
h({\bf R})+6$ results in identical spin configurations. The three 
nearest-neighbor vectors ${\bf e}_1$, ${\bf e}_2$ and ${\bf e}_3$ 
defined in Eq. (\ref{eq1}) are also displayed.} 
\label{fig1} 
\end{figure} 

\begin{figure}
\caption{Auto-correlation function of Fourier components of 
the height variables $z({\bf q})$ for $2S=1$.  
Only those corresponding to the five smallest
$|{\bf q}|^2$ are shown where we have averaged over data 
points with the same value of $|{\bf q}|^2$. The 
discrete ${\bf q}$-vectors comes about because of 
the periodic boundary condition used for the 
Monte Carlo simulation cell which consists of
$72\times 72$ spins. The solid lines in the figure 
are the fittings discussed in Sec. IV(B) to extrapolate 
the relaxation time $\tau_{\bf q}$ where we have used 
a cutoff in time $t\le 10$ measured in the unit of MSC.}  
\label{fig2} 
\end{figure} 

\begin{figure}
\caption{Relaxation time $\tau_{\bf q}$ as a function of 
${\bf q}^{-2}$ for $2S=1,...,8$. The solid lines are the 
fittings for cases of $2S=1,...,6$ from bottom to top 
(see Sec. V(B)). The dotted lines in the insert are only
the guide for eyes where data for $2S=6,7$ and $8$ are 
displayed from top to bottom.}  
\label{fig3} 
\end{figure} 

\begin{figure}
\caption{Structure factor $S_h({\bf q})$ of height variables. We show  
in the main figure the inverse of the structure factor 
as a function of ${\bf q}^2$ for $2S=1$. The lower 
envelope of the data corresponds to $q_y=0$. As an 
insert to the figure, we plot all the lower bounds for 
$2S=1,2,...,8$ which go from bottom to top. Solid lines 
in the insert are only the guide for eyes. Note that 
${\bf q}^2=17.5459...$ corresponds to the corner of the 
first Brillouin zone, i.e., ${\bf q}={\bf Q}\equiv 
{4\pi\over 3}(1,0)$.}  
\label{fig4} 
\end{figure} 

\begin{figure}
\caption{Scaling of $\langle |z({\bf q})|^2\rangle^{-1}$ as 
a function of ${\bf q}^2$. We have averaged data points 
of the same magnitude of ${\bf q}$-vector in each case of 
$2S=1,2,...,8$ obtained for system sizes $L=36, 48$, and $72$. 
Note the error bars are smaller than the symbol size.
Solid lines are fits using a cutoff ${\bf q}^2 \le 0.5$
discussed in Sec. V(C). Dotted lines are only guides for the eye.}    
\label{fig5} 
\end{figure}  

\begin{figure}
\caption{Extrapolation of stiffness constants. We show
$[{\bf q}^2 \langle |z({\bf q})|^2\rangle]^{-1}$
vs. ${\bf q}^2$ as log-linear plot
for $2S=1,2,...,6$. Note that we have performed an 
average over data points with the same magnitude of 
${\bf q}$-vector for each case of $2S$. Solid lines are 
the linear fitting discussed in Sec.~\ref{Stiffness} in order to extract 
the stiffness constant which is given by the intercept of the 
fitting. Also note that the fittings shown are performed with 
a cutoff ${\bf q}^2 \le 0.5$. Fittings with other cutoffs 
are discussed in the text.}  
\label{fig6} 
\end{figure}

\begin{figure}
\caption{Structure factor $S_E({\bf q})$ of the bond-energy operator near 
the zone corner ${\bf Q}$. Data points are averaged results 
over those with the same $|{\bf q-Q}|^2$ value for each case 
of $2S=1,2,...,8$.  
Note that data points for each $2S$ have been shifted upwards by 
$0.5$ with respect to their counterpart for $2S-1$ 
in order to disentangle the data. Solid lines are the 
fittings discussed in Sec. V(D) to extract the 
critical exponent $\eta_E$ of the bond-energy operator. 
Dotted lines 
are only to guide the eye.}
\label{fig7} 
\end{figure} 

\begin{figure}
\caption{Structure factor $S_{\sigma}({\bf q})$ of the spin operator 
near the zone corner ${\bf Q}$. Data points are averaged results
over those with the same $|{\bf q-Q}|^2$ value for each case
of $2S=1,2,...,8$.
Note that data points for each $2S$ are moved downwards by
$0.1$ with respect to their counterpart for $2S-1$
in order to disentangle the data. Solid lines are the
fittings discussed in Sec. V(D) to extract the
critical exponent $\eta_{\sigma}$ of the spin operator.
Dotted lines are only to guide the eye.}
\label{fig8}
\end{figure} 

\begin{figure}
\caption{Structure factor $S_{M}({\bf q})$ of the spin operator 
near the zone center ${\bf q}\to 0$. Data points are averaged results
over those with the same $|{\bf q}|^2$ value for each case
of $2S=1,2,...,8$. Note that data points for each $2S$ are 
moved upwards by $0.15$ with respect to their counterpart for $2S-1$
in order to disentangle the data. Solid lines are the
fittings discussed in Sec. V(D) to extract the critical exponent
$\eta_M$ of the uniform magnetization operator. Dotted lines are 
only to guide the eye.}
\label{fig9}
\end{figure}

\begin{figure}
\caption{Height distribution and ensemble average of the number of
free spins $n_s$, from one snapshot for each $S$ value.
On the top panel, we show the histograms of the 
height variables $h({\bf R})$ for $2S=2$ and $2S=8$. On the bottom 
panel, $n_s$ is displayed as a function of $2S$. Note that the maximum 
allowed value for $\langle n_s\rangle$ is $N_s/3$ where $N_s$ denotes 
the total number of spins in the simulation cell.} 
\label{fig10} 
\end{figure} 

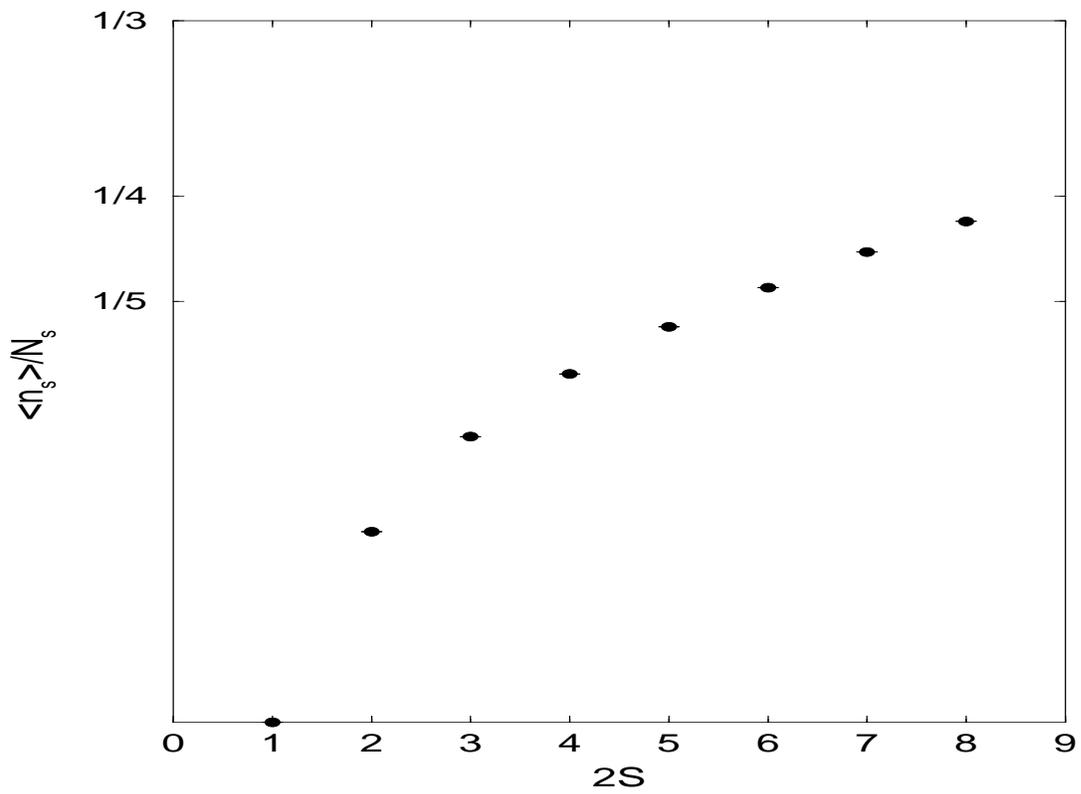
\end{document}